\documentclass[aps,prl,onecolumn,superscriptaddress]{revtex4}%
\usepackage{amsfonts}
\usepackage{amsmath}
\usepackage{amssymb}
\usepackage{graphicx}%
\begin{document}
\title{Topological Phases in the Single-Layer FeSe}
\author{Ningning Hao}
\affiliation{Beijing National Laboratory for Condensed Matter Physics and Institute of
Physics, Chinese Academy of Sciences, P. O. Box 603, Beijing 100190, China}
\affiliation{Department of Physics, Purdue University, West Lafayette, Indiana 47907, USA}
\author{Jiangping Hu}
\affiliation{Beijing National Laboratory for Condensed Matter Physics and Institute of
Physics, Chinese Academy of Sciences, P. O. Box 603, Beijing 100190, China}
\affiliation{Department of Physics, Purdue University, West Lafayette, Indiana 47907, USA}
\affiliation{Collaborative Innovation Center of Quantum Matter, Beijing, China}
\maketitle


\textbf{A distinct electronic structure was observed in the single-layer FeSe
which shows surprising high temperature superconductivity over 65k. Here we
demonstrate that the electronic structure can be explained by the strain
effect due to substrates. More importantly, we find that this electronic
structure can be tuned into robust topological phases from a topologically
trivial metallic phase by the spin-orbital interaction and couplings to
substrates. The topological phase is robust against any perturbations that
preserve the time-reversal symmetry. Our studies suggest that topological
phases and topologically related properties such as Majorana Fermions can be
realized in iron-based high T$_{c}$ superconductors. }

The single-layer (SL) FeSe
film\cite{Wang2012-fese,Liu2012-fese,He2013-fese,Tan2013-fese,Peng2014-fese,Zhang2014-fese,
Lee2014-fese} that is epitaxially grown on SrTiO$_{3}$(001) surface exhibits
several remarkable unique features compared with the bulk
FeSe\cite{Hsu2008-fese,Yeh2008-fese} and other Fe-based
superconductors\cite{Johnston2010-review,Dagotto2013-review}. In the bulk
FeSe, the superconducting transition temperature, $T_{c}\sim$%
8K\cite{Hsu2008-fese}, and the electronic structure is characterized by the
presence of both hole pockets around $\Gamma$ point and electron pockets
around M point in the first Brillouin zone[BZ]\cite{Subedi2008-dft,
Richard2011-review}. In the SL FeSe, $T_{c}$ exceeding 65K was
observed\cite{Wang2012-fese}. Furthermore, there are only electron pockets
around M point and the hole pockets at $\Gamma$ point are
absent\cite{Liu2012-fese,He2013-fese,Tan2013-fese}. The ARPES
experiments\cite{Liu2012-fese,He2013-fese} have shown that the electronic
structure of the SL FeSe cannot be obtained through the directly rigid band
shifting because a new band gap below Fermi surface is developed at M point.
The origin of the novel electronic structure and its relation to the high
$T_{c}$ in the SL FeSe remain as open issues.

\begin{figure}[pb]
\begin{center}
\includegraphics[width=1.0\linewidth]{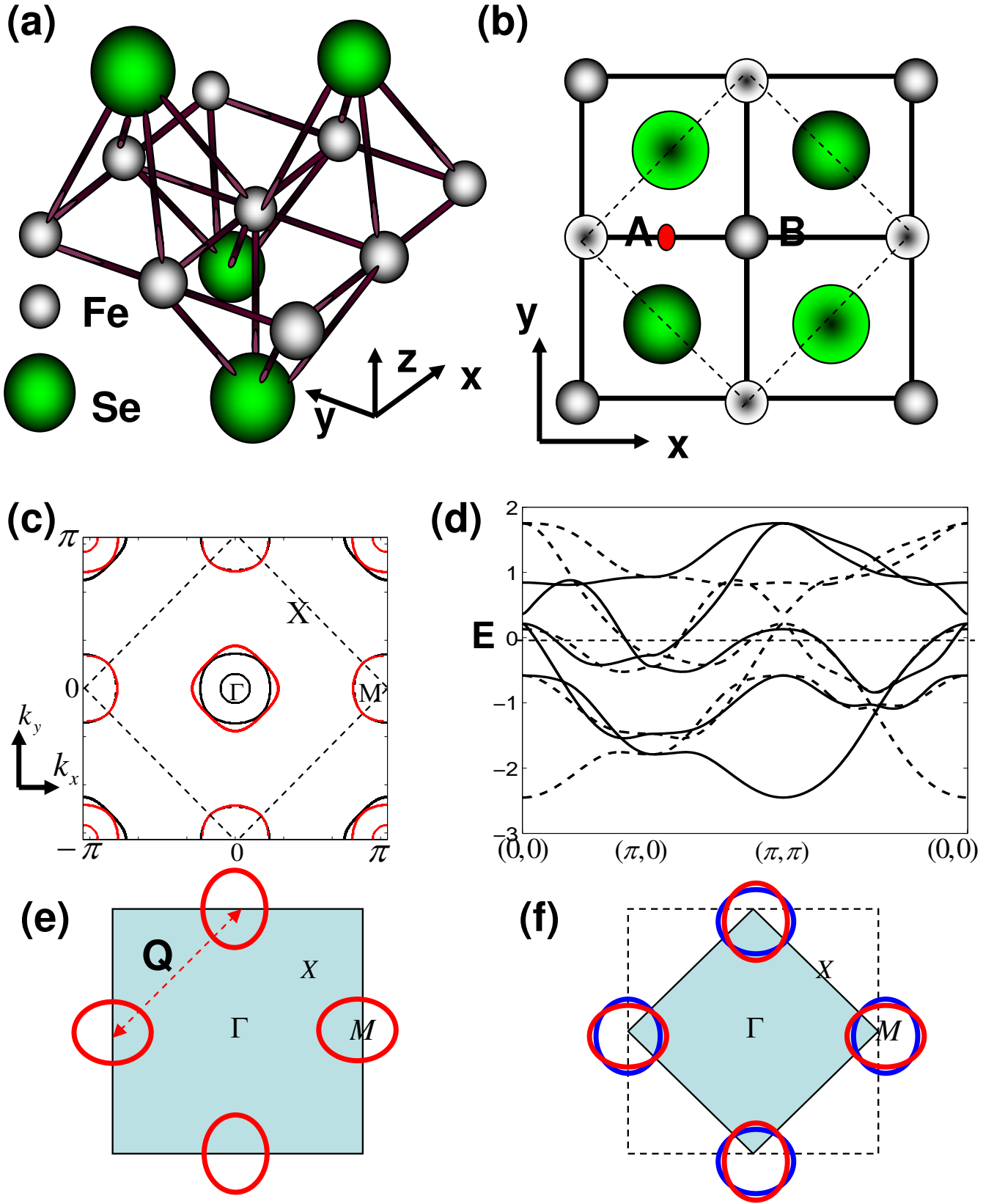}
\end{center}
\caption{(Color Online) (a) The lattice structure of FeSe. (b) The planform of
(a), the two-Fe unit cell is enclosed by the dashed lines with the two Fe
sublattice labeled with A and B. The inverse center is labeled by the red
elliptic spot at the mid point of A-B link. (c) the fermi surface for the bulk
FeSe, with three hole pockets locate around $\Gamma$ point and two electron
pockets around $M$ point in the two-Fe BZ. The two-Fe BZ is enclosed by the
dashed lines. (d) The band structure along high-symmetry lines. (e) and (f)
The schematic fermi surface for SL FeSe are shown in one Fe and two Fe unit
cell picture. The shadow region is the corresponding BZ. In (e), Q=($\pi$,
$\pi$) is the folding wave-vector. }%
\end{figure}


Here we first discuss the origin of the novel band structure as a result of
the lattice mismatch between the FeSe and substrate SrTiO$_{3}$. We
demonstrate that the lattice distortion can induce a phase transition around M
point from a gapless phase to a gaped phase and simultaneously suppress the
hole-like band at $\Gamma$ point.

It is known that the electronic structure in the bulk FeSe is rather
two-dimensional and is determined by the single FeSe tri-layer structure as
shown in Fig. 1(a) and (b) in which the unit cell includes two Fe labeled as
$A$ and $B$ and two Se. The band structure calculations from the density
functional theory show that only the five 3d orbitals of Fe play the
significant roles near Fermi surfaces\cite{Subedi2008-dft}. A general
effective d-orbitals model for the band structure in the real space can be
written as
\begin{equation}
H_{t}=\sum_{\alpha,\beta,\sigma}\sum_{mn}\sum_{ij}(t_{\alpha\beta,ij}%
^{mn}+\epsilon_{m}\delta_{mn}\delta_{\alpha\beta}\delta_{ij})d_{\alpha
,m,\sigma}^{\dag}(i)d_{\beta,n,\sigma}(j) \label{Ht_real}%
\end{equation}
Here, $\alpha,\beta$ label two different Fe A and B. $\sigma$ labels spin.
$m,n$ label five d orbitals: ($d_{xz}$, $d_{yz}$, $d_{x^{2}-y^{2}}$, $d_{xy}$,
$d_{z^{2}}$). $i,j$ label lattice sites. $t_{\alpha\beta,ij}^{mn}$ is the
corresponding hopping parameters. $\epsilon_{m}$ is the on-site energy of d
orbital. $d_{\alpha,m,\sigma}^{\dag}(i)$ creates an spin-$\sigma$ electron in
$m$th orbital of $\alpha$ Fe at site $i$. In the momentum space, the
Hamiltonian can be written as
\begin{equation}
H_{t}=\sum_{k}\phi^{\dag}(k)A(k)\phi(k)+\sum_{k^{\prime}}\phi^{\dag}%
(k^{\prime})A(k^{\prime})\phi(k^{\prime}) \label{Ht_momentum}%
\end{equation}
Here, $k$ is defined in the BZ of one Fe unit cell, and $k^{\prime}=k+Q$ with
$Q=(\pi,\pi)$. $\phi(k)=[d_{xz}(k),d_{yz}(k),d_{x^{2}-y^{2}}(k),d_{xy}%
(k),d_{z^{2}}(k)]^{T}$. $A(k)$ has been widely utilized to describe the
electronic structure of all kinds of iron-based
superconductors\cite{Kuroki2008-prl,Graser2009-njp}. The difference between
the one Fe unit cell and two Fe unit cell pictures can be found in section I
of supplementary materials.

The Fermi surface and band structure along high-symmetry lines are shown in
Fig. 1(c) and (d). In Fig. 1(c), there are three hole pockets around the
$\Gamma$ point and the two electron pockets around $M$ point. We have
specified that the black pockets in Fig.1(c) are from $A(k)$ while the red
pockets from $A(k^{\prime})$. In Fig. 1(d), the solid or dashed lines denote
the bands from $A(k)$ or $A(k^{\prime})$.

\begin{figure}[pt]
\begin{center}
\includegraphics[width=1.0\linewidth]{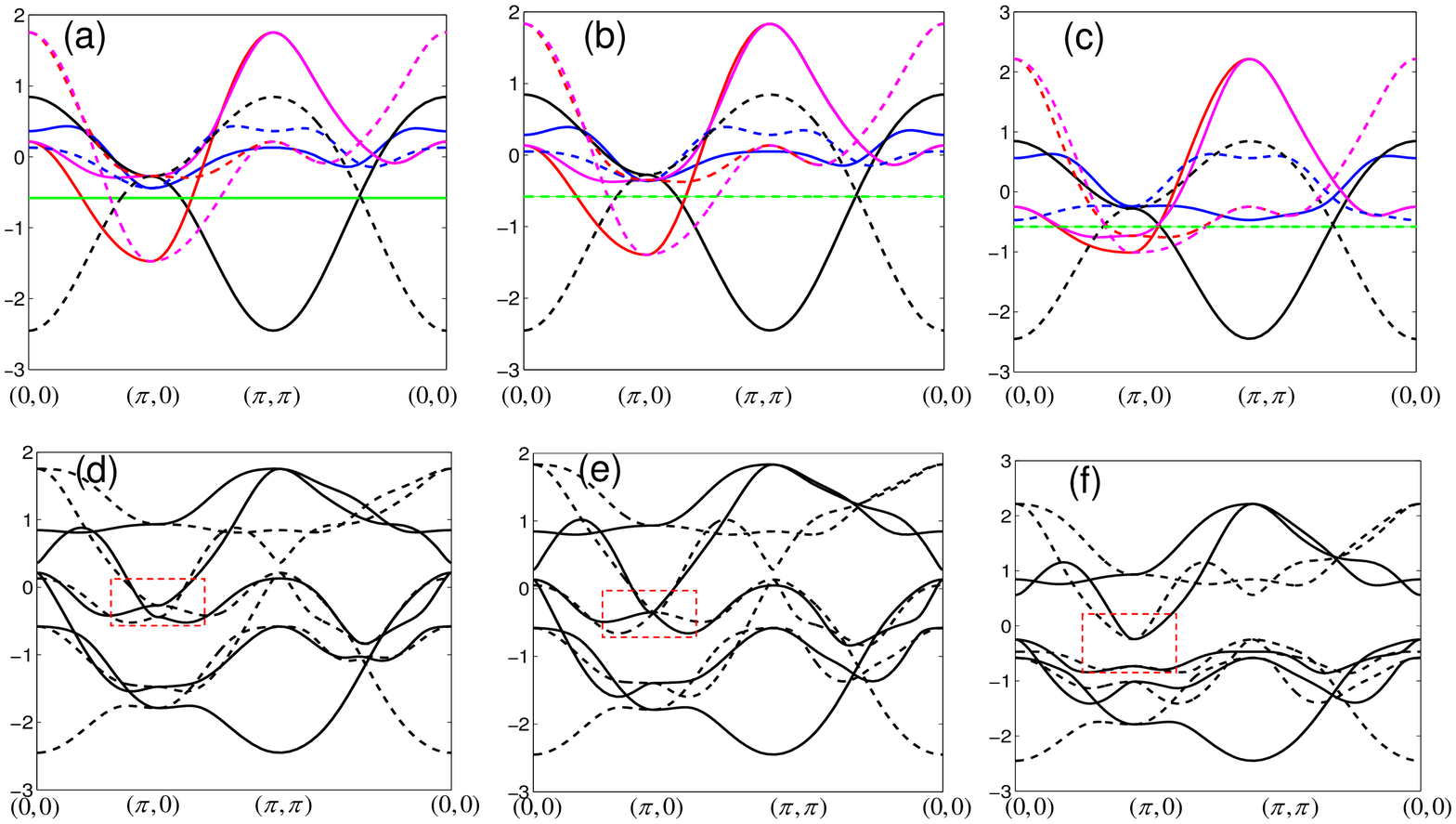}
\end{center}
\caption{(Color Online) (a)-(c) The bands structures along high-symmetry line
for the case without interorbital hoppings. (a) for the bulk case. (b) the
adjusted parameters are $t_{xy}^{44}=0.066$, $t_{x}^{14}=0.405$, $t_{x}%
^{11}=-0.120$. (c) $t_{xy}^{44}=0.036$, $t_{x}^{44}=0.163$, $t_{x}^{14}%
=0.405$, $t_{x}^{11}=-0.311$. The solid/dashed lines are from $A(k)$/$A(k+Q)$.
The red/magenta/blue/black/green lines are the $d_{xz}$/$d_{yz}$/$d_{xy}%
$/$d_{x^{2}-y^{2}}$/$d_{z^{2}}$ bands. (d)-(f) are the cases with
inter-orbital hoppings and correspond to (a)-(c). }%
\end{figure}

In Fig. 1, there is no gap openning in the band structure at M point. To gain
the insight of the gap openning observed experimentally in the SL FeSe, we
analyze the symmetry characters of the bands. As shown in Fig. 1, there are
four bands along $\Gamma$-M direction near Fermi surfaces. Each $A(k)$ and
$A(k^{\prime})$ contribute two bands. The bands contributed from $A(k)$ and
$A(k^{\prime})$ have opposite parity\cite{Hu2013-odd,Hu2012-s4}. The two bands
from $A(k)$ belong to $A_{2}$ and $B_{2}$
representations\cite{Cvetkovic2013-band}, which is the reason why the two
bands can cross each other without showing the sign of hybridization. However,
the two bands from $A(k^{\prime})$ belong to the same $B_{1}$
representations\cite{Cvetkovic2013-band}. If these two bands across each
other, the hybridization must take place. We notice that these two bands are
attributed to the $d_{xz}$ and $d_{xy}$ orbitals respectively near M point.
Moreover, the symmetry characters of the bands do not depend on the
inter-orbital couplings because the couplings between two different d-orbitals
at high symmetry points, such as $\Gamma$ and $M$, vanish. Therefore, we can
adjust the relative energy difference between $d_{xz}$ and $d_{xy}$ at M
points to create an crossing between the two bands. such an crossing can
result in a gap openning at M point because of the hybridization.

To confirm above symmetric analysis, we present an evolution of the band
structure by manipulating hopping parameters in Fig.2. The bulk band
structures are shown in Fig.2 (a) and (d). Note that we set all inter-orbital
hoppings zero in Fig.2 (a) in order to clearly distinguish the five
$d$-orbital bands. In Fig. 2 (a), the red-dashed $d_{xz}$ band and the
blue-dashed $d_{xy}$ band has no cross along (0,0)-($\pi,0$) line. In Fig. 2
(d), there is no band gap opening around M point, i.e., ($\pi,0$)( see the red
rectangle region in Fig. 2(d)). When we adjust the values of some hopping
parameters (See the caption of Fig. 2), the red-dashed $d_{xz}$ band is pushed
down and the blue-dashed $d_{xy}$ band is pushed up. The critical case is
shown in Fig. 2 (b), the two bands meet each other at $M$ point. Then, the
continued push leads to the cross between red-dashed $d_{xz}$ and blue-dashed
$d_{xy}$ bands at some point between (0,0) and ($\pi,0$) (See Fig. 2 (c)).
Turning on the inter-orbital hoppings, we obtain a new band structure shown in
Fig. 2 (f). Compared with the bulk bands in Fig. 2(d), the bands in Fig. 2(f)
present some new feature, such as (1) the hole pockets around $\Gamma$ point
are pushed down the fermi energy with only electron pockets around $M$ point
surviving. (2) the band top of hole pockets nearly has same energy as the band
bottom of electron pockets. (3) there is a band gap opened at $M$ point (See
the red rectangle region Fig. 2 (f)). These features of band structure shown
in Fig. 2 (f) are comparable to the ARPES
observations\cite{Liu2012-fese,He2013-fese}.

From Fig. 2, we can find that the evolution of band structure from bulk FeSe
to a SL FeSe is strongly sensitive to the change of a single hopping parameter
$t_{x}^{11}$, i.e., the amplitude of intra-$d_{xz}/d_{yz}$ hopping along $x/y$
directions. Namely, the nearest-neighbor intra-orbital hoppings for $d_{xz}$
and $d_{yz}$ changes from the strong anisotropy in the bulk FeSe to near
isotropy in the SL FeSe. (See the Fig.7 in supplementary materials). We argue
that this is the essential reason to drive the electronic structure of the SL
FeSe. The SL FeSe film is epitaxially grown on SrTiO$_{3}$(001) surface.
Consequently, the lattice constant for the SL FeSe film should match the
lattice constant for SrTiO$_{3}$, i.e., 3.905\AA . Compared with the bulk FeSe
with lattice constant 3.76\AA \cite{Hsu2008-fese}, the apparent lattice
mismatch between FeSe and SrTiO3 should exert a strong tensile strain on the
FeSe film, and drive the lattice distortion for the FeSe
film\cite{Tan2013-fese,Peng2014-fese}. The lattice distortion naturally
induces the changes of hopping. The influence of lattice distortion to the
hopping parameters is discussed in details in the section II of supplementary materials.

After obtaining the new band structure, we can ask whether the new band gap at
$M$ point (See the red rectangle region Fig. 2 (f)) is topologically trivial
or nontrivial when the spin-orbital coupling (SOC) is considered? As the
$t_{2g}$ orbitals are significant near Fermi surface, we can write SOC within
the $t_{2g}$ subset. Up to the next nearest neighbor, the general SOC
Hamiltonian in momentum space can be written as

%

\begin{align}
H_{so}  &  =H_{so1}+H_{so2},\label{Hso}\\
H_{so1}  &  =\sum_{\tilde{k}=k,k^{\prime}}(-1)^{\sigma}\lambda_{\perp}%
(\tilde{k})d_{xz,\sigma}^{\dag}(\tilde{k})d_{yz,\sigma}(\tilde{k}%
)+H.c.,\label{Hso1}\\
H_{so2}  &  =\sum_{\tilde{k}=k,k^{\prime}}[i\lambda_{\shortparallel,xz}%
(\tilde{k})d_{xz,\uparrow}^{\dag}(\tilde{k})+\lambda_{\shortparallel
,yz}(\tilde{k})d_{yz,\uparrow}^{\dag}(\tilde{k})]d_{xy,\downarrow}(\tilde
{k}+Q)+H.c.\nonumber\\
&  -\sum_{\tilde{k}=k,k^{\prime}}d_{xy,\uparrow}^{\dag}(\tilde{k}%
)[i\lambda_{\shortparallel,xz}(\tilde{k})d_{xz,\downarrow}(\tilde
{k}+Q)+\lambda_{\shortparallel,yz}(\tilde{k})d_{yz,\downarrow}(\tilde
{k}+Q)]+H.c. \label{Hso2}%
\end{align}
Here, $\sigma=\mp$ for spin $\uparrow$ or $\downarrow$. $\lambda_{\perp
}(\tilde{k})=(\lambda_{\perp}^{o}+4\lambda_{\perp}^{nn}\cos\tilde{k}_{x}%
\cos\tilde{k}_{y})$. $\lambda_{\shortparallel,xz/yz}(\tilde{k})=\lambda
_{\shortparallel}^{o}+2\lambda_{\shortparallel}^{n}\cos\tilde{k}_{x/y}$. The
index $\alpha=o,n,nn$ in the $\lambda_{\shortparallel,\perp}^{\alpha}$
indicate the onsite, nearest neighbor, and next nearest neighbor SOC
respectively. $H_{so1}$ describes the SOC between $d_{xz}$ and $d_{yz}$
orbitals. This term does not flip spin and also conserves momentum with
respect to the 1-Fe unit cell. $H_{so2}$ describes the SOC between $d_{xy}$
and $d_{xz,yz}$. This term flips spin and does not conserve momentum with
respect to the 1-Fe unit cell. Therefore, $H_{so2}$ essentially break the
non-symmorphic lattice symmetry in the SL FeSe, namely, an one-unit
translation along the Fe-Fe direction followed by a mirror reflection with
respect to the layer\cite{Lee2008-iron}. In the presence of $H_{so2}$, the
2-Fe unit cell can not be reduction to the 1-Fe unit cell.

The $H_{so2}$ is not the only term that breaks the reduction to the 1-Fe unit
cell. If we consider the effect of a substrate, the space inversion symmetry
is naturally broken for the SL FeSe. Such a parity breaking can also result in
a term that only preserves the 2-Fe unit cell\cite{Hu2013-odd},
\begin{equation}
H_{s}=\sum_{m,\sigma}\sum_{\tilde{k}=k,k^{\prime}}\xi_{s}(\tilde
{k})d_{m,\sigma}^{\dag}(\tilde{k})d_{m,\sigma}(\tilde{k}+Q).
\label{Hpmomentum}%
\end{equation}
Here, $\xi_{s}(\tilde{k})$ measures this parity breaking effect. Likewise, we
only focus on three $t_{2g}$ orbitals. If we take a constant $\xi_{s}%
(\tilde{k})$, the term describes a staggered potential in the iron square lattice.

The full Hamiltonian for the electronic structure of the SL FeSe with a
substrate is a combination of $H_{t}$, $H_{so}$ and $H_{s}$, namely,%
\begin{equation}
H=H_{t}+H_{so1}+ H_{so2}+H_{s}. \label{Htot}%
\end{equation}
The last two terms in $H$ only preserve the translational symmetry with
respect to the 2-Fe unit cell.

We first concentrate on the effect of $H_{so1}$ which preserves the 1-Fe unit
cell. Namely, we ignore both $H_{so2}$ and $H_{s}$. The glancing features can
be spied through the numerical results of the band structure of the
Hamiltonian $H=H_{t}+H_{so1}$. The results are shown in Fig. 3. In Fig. 3
(a)-(c), it is explicitly shown that the band gap undergoes a closing and
reopening process at $M$ point when $\lambda_{\perp}^{o}$ and $\lambda_{\perp
}^{nn}$ are tuned from zero to some finite values. Such a phenomena is a
strong indication of a topological phase
transition\cite{Kane2005-qsh,Bernevig2006-ti}. Fig. 3 (d)-(i) provide a clear
proof of the occurrence of the topological phase transition. The spectra in
Fig. 3(d)-(f) and Fig.3 (g)-(i) are plotted within the 1-Fe unit cell and the
2-Fe unit cell respectively. The latter can be obtained through folding the
1-Fe unit cell picture with the folding wave vector $Q=(\pi,\pi)$. It is clear
that the signature of the topological phase, gapless edge states, emerges
after the band gap reopens.

\begin{figure}[pt]
\begin{center}
\includegraphics[width=1.0\linewidth]{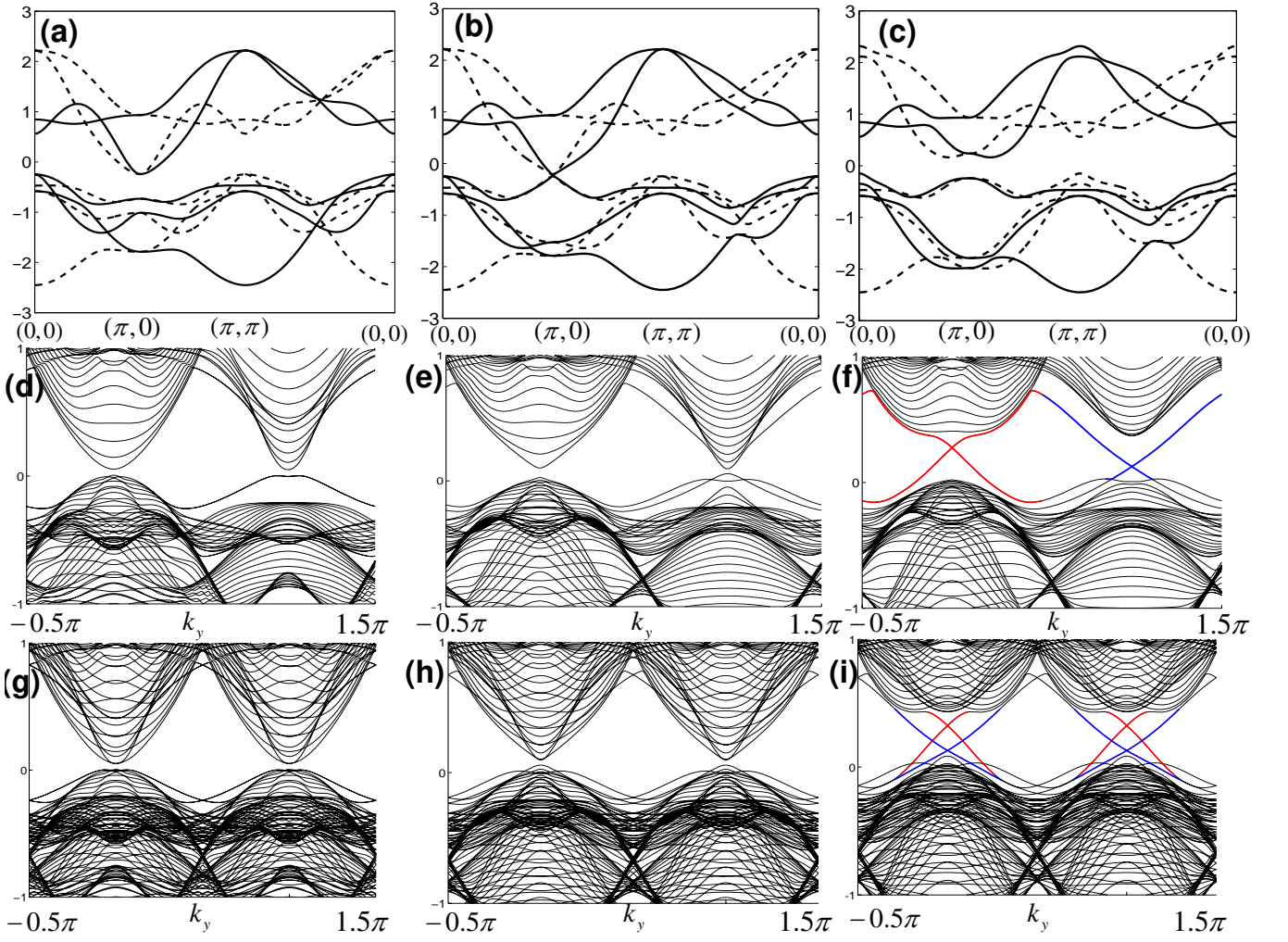}
\end{center}
\caption{(Color online) (a) (b) (c) The spectrum along high-symmetry line for
different spin-orbital coupling parameters with (a) $\lambda_{\perp}^{o}=0$,
$\lambda_{\perp}^{nn}=0$ (b) $\lambda_{\perp}^{o}=0.32$, $\lambda_{\perp}%
^{nn}=0.08$ (c) $\lambda_{\perp}^{o}=0.5$, $\lambda_{\perp}^{nn}=0.12$. Other
parameters $\lambda_{\shortparallel}^{o}$, $\lambda_{\shortparallel}^{n}$ and
$\xi_{s}$ are set zero. (d) (e) (f) ( (g) (h) (i)) are the corresponding one
Fe unit cell (two Fe unit cell) edge spectrum with open boundary along $x$
direction. the width of the single-layer film is 21 in unit of lattice
constant. }%
\end{figure}

We can also construct an effective Hamiltonian to describe the topological
transition. The detailed derivation is presented in the section III of
supplementary materials. Here, we only briefly summarize the result. We define
the new basis for $d_{xz}$ and $d_{yz}$ orbitals according to the eigenstates
of azimuthal quantum number $l=2$ and magnetic quantum number $m=\pm1$ i.e.
$d_{(2,1),\sigma}(\tilde{k})=-\frac{1}{\sqrt{2}}[d_{xz,\sigma}(\tilde
{k})+id_{yz,\sigma}(\tilde{k})]$, $d_{(2,-1),\sigma}(\tilde{k})=\frac{1}%
{\sqrt{2}}[d_{xz,\sigma}(\tilde{k})-id_{yz,\sigma}(\tilde{k})]$.
Around each $M$ point, the band structure can be spanned by $\Psi_{eff}%
(\tilde{k})=[\phi_{eff,\uparrow}(\tilde{k}),\phi_{eff,\downarrow}(\tilde
{k})]^{T}$ with $\phi_{eff,\sigma}(\tilde{k})=[d_{xy,\sigma}(\tilde
{k}),d_{(2,-(-1)^{\sigma}),\sigma}(\tilde{k})]^{T}$ and is described by the
effective Hamiltonian,
\begin{equation}
H_{eff}=\sum_{\tilde{k}=k,k^{\prime}}\Psi_{eff}^{\dag}(\tilde{k}%
)A_{eff}(\tilde{k})\Psi_{eff}(\tilde{k}) \label{Htoteff1}%
\end{equation}
where,
\begin{equation}
A_{eff}(\tilde{k})=\sum_{a=0}^{5}\varepsilon_{a}(\tilde{k})\Gamma^{a}%
+\sum_{a<b=1}^{5}\varepsilon_{ab}(\tilde{k})\Gamma^{ab} \label{Heffnew}%
\end{equation}
is a $4\times4$ matrix. The $\Gamma$ matrices are defined as $\Gamma
^{(0,1,2,3,4,5)}=(\tau^{0}\otimes s^{0},\tau^{0}\otimes s^{z},\tau^{0}\otimes
s^{y},\tau^{z}\otimes s^{x},\tau^{y}\otimes s^{x},\tau^{x}\otimes s^{x})$,
where the Pauli matrices $\tau$ and $s$ span the orbital and spin subspaces.
$\Gamma^{ab}=[\Gamma^{a},\Gamma^{b}]/(2i)$. The non-zero elements are
$\varepsilon_{0/1}(\tilde{k})=\frac{1}{2}[E_{xy}(\tilde{k})\pm E_{(2,-1)}%
(\tilde{k})\pm\left\vert \lambda_{\perp}(\tilde{k})\right\vert ]$ and
$\varepsilon_{12/13}(\tilde{k})=\pm\sqrt{2}t_{x}^{14}\sin\tilde{k}_{y/x}$, in
which $E_{xy}(\tilde{k})=A_{44}(\tilde{k})$ and $E_{(2,-1)}(\tilde{k}%
)=\frac{1}{2}[A_{11}(\tilde{k})+A_{22}(\tilde{k})]$.

The above effective Hamiltonian has the same form as those for HgTe quantum
wells\cite{Bernevig2006-ti}. At each $M$ point, the band gap of Eq.
(\ref{Heffnew}) is measured by a Dirac mass $\varepsilon_{1}(\tilde{k})$. When
the $\lambda_{\perp}(\tilde{k})$ overcomes the trivial band gap measured by
$E_{xy}(\tilde{k})-E_{(2,-1)}(\tilde{k})$, namely, the mass $\varepsilon
_{1}(\tilde{k})<0$ changes its sign from positive to negative, a trivial to
nontrivial topological phase transition takes place.

\begin{figure}[pt]
\begin{center}
\includegraphics[width=1.0\linewidth]{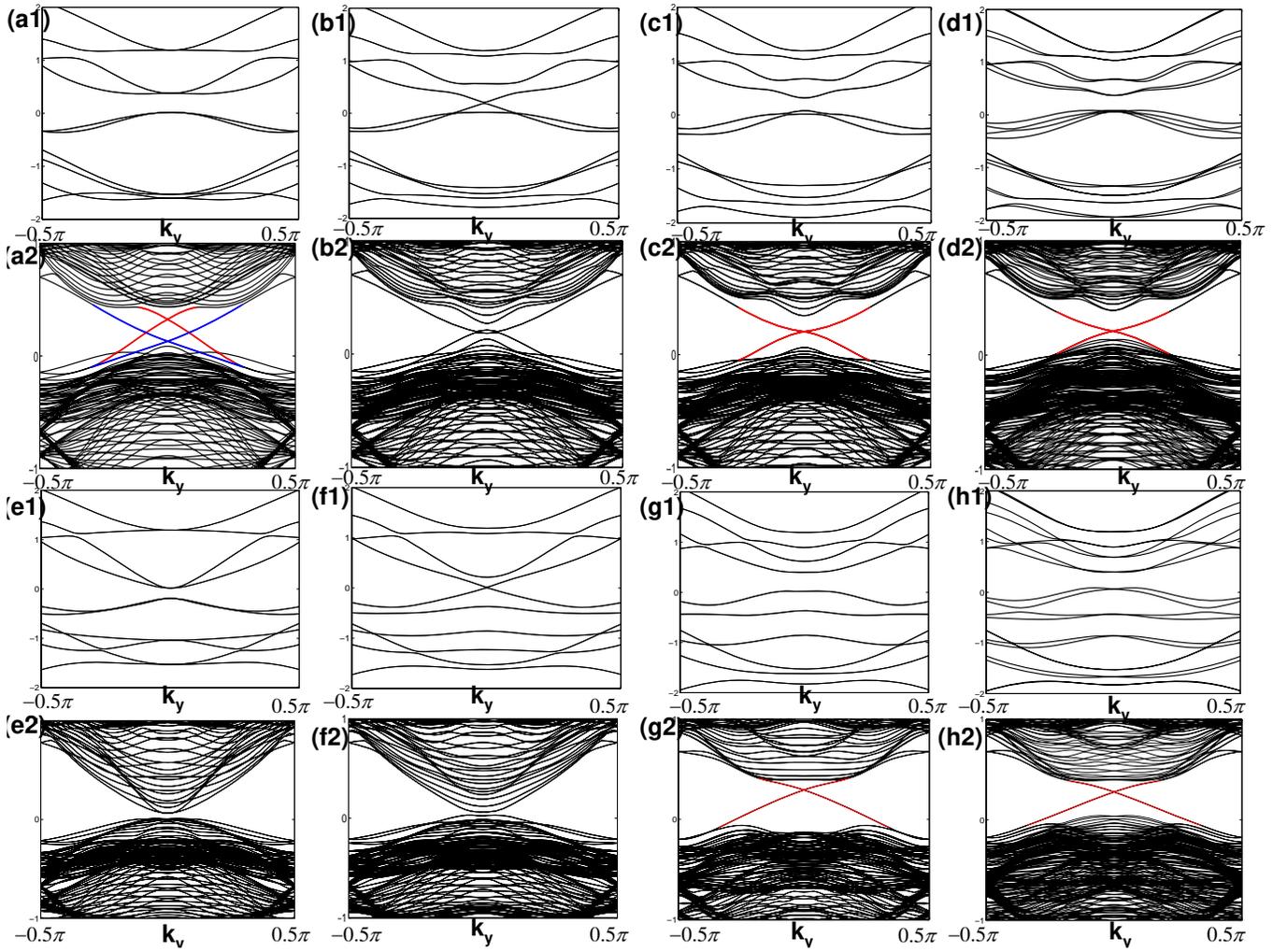}
\end{center}
\caption{(Color online) (a1)-(d1) The evolution of spectrum with $k_{x}=\pi$,
$k_{y}\in\lbrack-\frac{\pi}{2},\frac{\pi}{2}]$ for parameters ($\lambda
_{\perp}^{o},\lambda_{\perp}^{nn},\lambda_{\shortparallel}^{o},\lambda
_{\shortparallel}^{n},\xi_{s}$) with the values (0.5,-0.12,0,0,0) in (a1),
(0.5,-0.12,0,0,0.19) in (b1), (0.5,-0.12,0,0,0.3) in (c1) and
(0.5,-0.12,0.2,0.3) in (d1). (a2)-(d2) are the corresponding edge spectrum
with open boundary along x direction. (a2) is weak topological phase (c2) and
(d2) are strong topological phase. (b2) is the critical point. (e1)-(h1) The
evolution of spectrum with $k_{x}=\pi$, $k_{y}\in\lbrack-\frac{\pi}{2}%
,\frac{\pi}{2}]$ for parameters ($\lambda_{\perp}^{o},\lambda_{\perp}%
^{nn},\lambda_{\shortparallel}^{o},\lambda_{\shortparallel}^{n},\xi_{s}$) with
the values (0.2,-0.05,0,0,0) in (e1), (0.2,-0.05,0,0,0.2) in (f1),
(0.2,-0.05,0,0,0.6) in (g1) and (0.2,-0.05,0.2,0.6) in (h1). (e2)-(h2) are the
corresponding edge spectrum with open boundary along x direction. (e2) is
trivial phase (g2) and (h2) are strong topological phase. (f2) is the critical
point. In (a2)-(d2) and (e2)-h(2), the width of the single-layer film is 21
unit of Fe-Fe lattice constant. }%
\end{figure}We notice that there are even number (two) nontrivial Dirac cone
structures in the Hamiltonian $H=H_{t}+H_{so1}$ as shown in Fig. 3. This is
because that the Hamiltonian preserves the full non-symmorphic lattice
symmetry so that the band structure decouples into two independent parts in
the view of the 2-Fe unit cell. In the 2-Fe unit cell BZ, there is a
nontrivial Dirac cone structure at the BZ zone corner for each part. If we
understand this in the 1-Fe unit cell picture, a nontrivial Dirac cone
structure exists at each $M$ point of the 1-Fe BZ. As each Dirac cone
structure results in a pair of edge states in a $Z_{2}$ time reversal
invariant topological phase\cite{Kane2005-qsh, Bernevig2006-ti}, there are two
pairs of edge states in this topological phases as shown in Fig. 3. With even
pairs of edge states, the topological phase essentially is unstable. One can
imagine that if any perturbations that break non-symmorphic lattice symmetry
may lead to a coupling between two pairs of edge states to open a gap on edge
states. Therefore, we call this topological phase as the weak topological
phase\cite{Fu2007-prb}.

As we mentioned earlier, in the general Hamiltonian (Eq.\ref{Htot}) for the SL
FeSe with a substrate, both $H_{so2}$ and $H_{s}$ break the non-symmorhpic
lattice symmetry. Therefore, we have to answer how these two terms affect the
weak topological phase. First, we ask the effect of $H_{so2}$. Indeed, the
spin-flip term in $H_{so2}$ causes the couplings between the two Dirac cones
to create a gap on the edge states. If we assume that the SOC parameter
$\lambda_{\shortparallel}^{o}$ is small, the gap opened on the edge states is
given by $\sim\left\vert \lambda_{\shortparallel}^{o}\right\vert
^{2}/\left\vert t_{x}^{14}\right\vert $. Therefore, in principle, the
$\lambda_{\shortparallel}^{o}$ term in Eq. (\ref{Hso2}) can be considered as a
controlling parameter of the gap. The situation is very similar to a
topological crystalline insulator\cite{Fu2011-tci,Hseih2012-ti} and the
topological phase in a system with a non-symmorphic lattice
symmetry\cite{Liu2013-ns}.

\begin{figure}[pt]
\begin{center}
\includegraphics[width=1.0\linewidth]{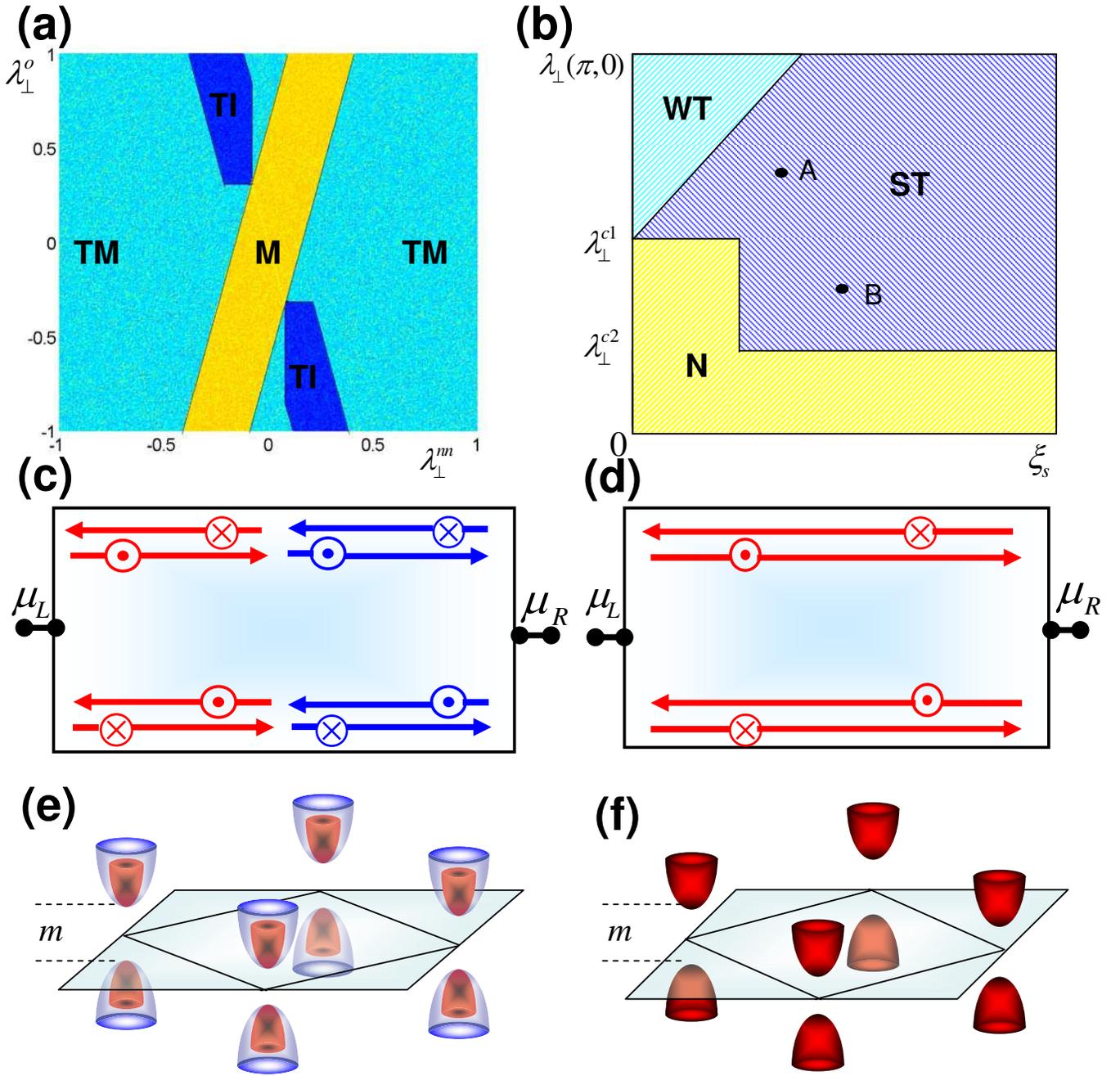}
\end{center}
\caption{(Color online) (a) The phase diagram as a function of $\lambda
_{\perp}^{o}$ and $\lambda_{\perp}^{nn}$. Here, M, TM and TI label metal,
topological metallic and topological insulating phases, respectively. (b) The
phase diagram as a function of $\lambda_{\perp}(\pi,0)$ and $\xi_{s}$ for
$\lambda_{\perp}(\pi,0)>0$. Here, WI, ST and N label weak topological, strong
topological and normal trivial phases, respectively. (c) and (d) The picture
of the accumulation of spin currents in the weak and strong topological
insulating phases of SL FeSe with the rectangle geometry. Here, "$\otimes$",
"$\odot$" label majority spin-up and -down. The color corresponds to Fig. 3(i)
and Fig. 4(d2) and (h2). (e) The two non-trivial (trivial) Dirac cones with
\ negative (positive) mass $m$ correspond to Fig. 4 (a1) (Fig.4 (e1)) around
$M$ points in BZ. The red and blue color labels the Dirac cone structures with
even and odd parities. (f) The single non-trivial Dirac cone with negative
mass $m$ (Fig. 4 (c1) and (g1)) around $M$ points in BZ. }%
\end{figure}

However, if we turn on $H_{s}$, the situation is drastically different. Rather
than destroying the topological phase, we find amazingly that $H_{s}$ can
stabilize the topological phase and drives the system to a strong topological
phase. To understand it, we consider the effective Hamiltonian in
Eq.\ref{Htoteff1} describing the weak topological phase. If we add $H_{s}$,
for $\lambda_{\perp}(k)>0$, the spectrum becomes $E(k)=\pm\sqrt{\varepsilon
_{12}^{2}(k)+\varepsilon_{13}^{2}(k)+[\varepsilon_{1}(k)\pm\xi_{s}]^{2}}$
where we have set effective long-range $\lambda_{\shortparallel}^{n}=0$ for
simplicity. We can find that the effect of $H_{s}$ is to change the Dirac
masses at the two M points. The changes, $\varepsilon_{1}(k)\pm\xi_{s}$, are
different for the Dirac cones at the two different $M$ points. Therefore,
$H_{s}$ can create a band inversion only in one Dirac cone but not the other,
a case for a strong topological phase with odd number of non-trivial Dirac
cones. Thus, the strong topological phase is robust against any
non-time-reverse-symmetry broken couplings, including the $H_{so2}$, as long
as the coupling does not close the bulk energy gap.

The numerical proof of the above analysis is plotted in Fig. 4 in which the
edge spectra in Fig. 4(d2) and (h2) clearly indicate a strong topological
phase. More specifically, we discuss the strong topological phases in two
cases: (1) $\varepsilon_{1}(k)<0$ and (2) $\varepsilon_{1}(k)>0$. In the first
case, when $\left\vert \varepsilon_{1}(k)\right\vert <\xi_{s}$, one of
non-trivial Dirac cones undergoes another gap-close-and-reopen process, and
becomes trivial one with positive mass $\varepsilon_{1}(k)+$ $\xi_{s}>0$. At
each $M$ point, only one non-trivial Dirac cone survives. The band evolution
for this process is shown in Fig. 4(a1)-(d2). In the second case, when
$\left\vert \varepsilon_{1}(k)\right\vert <\xi_{s}$, one of trivial Dirac
cones undergoes a gap-close-and-reopen process, and becomes a non-trivial one
with negative mass $\varepsilon_{1}(k)-$ $\xi_{s}<0$. At each $M$ point, only
one non-trivial Dirac cone emerges. The band evolution for this process is
shown in Fig. 4(e1)-(h2). In the first case, one needs relatively large SOC
and the parity broken coupling to overcome the trivial band gap and eliminate
one non-trivial Dirac cone. In the second case, we can see that a finite value
$H_{s}$ can create a region of a strong topological phase and can dramatically
reduce the critical SOC that is necessary for a topological phase.

The phase diagrams for the topological transition are plotted in Fig. 5. In
Fig. 5 (a), we plot the phase diagram for the case of the weak topological
phase with respect to the parameters, $\lambda_{\perp}^{o}$ and $\lambda
_{\perp}^{nn}$. Here, the topological metal phase means the hole-like band top
at $\Gamma$ point is higher than the electron-like band bottom around $M$
point so that we cannot tune the system into a full insulating phase, but
there is a gap at $M$ point to protect the topological phase. In the
topological insulating phase, the edge states propagate along the edges of the
materials with opposite speed for different spin components. The pictures of
the edge states and the Dirac cones is schematically shown in Fig. 5 (c) and
(e) respectively. In Fig 5.(b) we plot the phase digram for $\lambda_{\perp
}(k)>0$ case with respect to the parameter $\xi_{s}$ in $H_{s}$ and
$\lambda_{\perp}(\pi,0)$. The A and B points correspond to the cases shown in
Fig. 4. (c1) and (g1). The transport picture for the edge states is shown in
Fig. 5 (d), which matches the single nontrivial Dirac cone structure shown in
Fig. 5(f).

According to the aforementioned discussion, we see that the topological phase
is associated to three key parameters: the trivial band gap, the intrinsic
spin-orbital coupling strength at M points and the parity broken coupling
induced by the substrates. In the FeSe, the bare spin-orbital coupling
strength of Fe atoms is around 80$mev$\cite{Tiago2006-so}, which can cause a
bare energy splitting $\sim$ 100 $mev$ around $M$ point. Considering the renormalization effect (about 4 in FeSe), the real splitting of the band caused by the SOC would be around 25mev.
 Although we may
replace Fe by heavier atoms such as Ru to further increase the SOC strength,
the strength of SOC, more or less, is a fixed quantity in the FeSe. However,
the trivial band gap and parity-broken coupling at M point can be engineered.
With different substrates, the in-plane lattice constant can be tuned from
3.67 to 4$\mathring{A}$, which has been recently demonstrated in
FeSe/STO/KTO\cite{Peng2014-prl} and FeSe/BTO\cite{Peng2014-fese} structures.
There are a zoo of substrates to play for this interacial
engineering\cite{Peng2014-fese}. In the supplement, we have shown the
qualitative relationship between the gap and lattice parameters. An accurate
quantitative prediction of the trivial band gap on different substrates are
beyond the capability of any current numerical methods. Nevertheless, as we
know that in one limit, the large band gap (about 50mev measured by ARPES\cite{Peng2014-fese}) is created in the SL FeSe with STO and
BTO substrates, and in the other limit, it vanishes in the bulk FeSe, we can
be sure that the interracial engineering technique can create a tunable gap in
a SL FeSe. Furthermore, the parity-broken coupling can be larger when the
interaction to a substrate is stronger. $\xi_{s}$ is tunable with large
flexibility. Hence, realizing the topological phase in the SL FeSe is very promising.

It is also possible to further extend above analysis to the bulk materials.
According to our above discussions, the difference of the electronic
structures between the SL FeSe and the bulk FeSe mainly originates from the
lattice distortion induced by the substrate, and the ratio between the height
of Se to the Fe plane and the length of Fe-Fe bond uniquely measures this
difference. When the ratio declines less than the threshold value, a band gap
is opened around $M$ point. Experimentally, this ratio can be tuned through
applying internal or external pressure to the materials. Therefore, we suggest that this
topological transition may be realized in the bulk materials of iron-based
superconductors if the intensity of SOC is comparable with the band gap around
M point.

We also want to emphasize a great advantage to realize the topological phase
in iron-based superconductors. Unlike the conventional hybridized system
proposed to realize topological superconductors\cite{Qi2011-review} and
Majorana Fermion\cite{Fu2008-ts,Lutchyn2010-mf} in which the superconductivity
is induced through the proximity effect of conventional s-wave superconductor
with the low $T_{c}$, the iron-based materials present high-$T_{c}$
superconductivity themselves, especially, the SL FeSe shows superconductivity
at very high temperature. Hence, a significant importance of this study is to
provide a realistic possibility to realize the stable Majorana Fermions at
high temperature.

In conclusion, we show that the single layer FeSe present distinct and
abundant structures compared with the bulk FeSe through the interaction with
the substrates. We predict that there exists strong topological phase in the
SL FeSe. It is conceivable that many important physics and applications can
emerge when nontrivial topology meets the high-$T_{c}$ superconductivity.

The work is supported by the Ministry of Science and Technology of China 973
program(2012CB821400) and NSFC.


\end{document}